\pdfoutput=1 
\documentclass{JINST}

\title{CMS RPC muon detector performance with 2010-2012 LHC data}

\author{G. Pugliese$^a$,
 {\rm on behalf of the CMS Collaboration}\\
\llap{$^a$}Dipartimento Interateneo di Fisica and Sezione INFN, Via G. Amendola 173, 70126 Bari, Italy ,\\
E-mail: \email{gabriella.pugliese@ba.infn.it}}

\abstract{The  muon spectrometer  of the CMS  (Compact Muon Solenoid) experiment at the Large Hadron Collider (LHC) is equipped with a redundant system made of Resistive Plate Chambers and Drift Tube in barrel and RPC and Cathode Strip Chamber in endcap region. In this paper, the operations and performance of the RPC system during the first three years of LHC activity will be reported. The stability of RPC performance,  such as efficiency, cluster size and noise, will be reported. 
Finally, the radiation background levels on the RPC system have been measured as a function of the LHC luminosity. Extrapolations to the LHC and High Luminosity LHC conditions  are also discussed.}


\begin{document}

\section{Introduction} 

Compact Muon Solenoid (CMS) \cite{cms} is a multi-purpose particle physics detector built at the CERN Large Hadron 
 Collider (LHC) \cite{lhc}.  The CMS basic design concept is a super-conducting solenoid that allows a magnetic field of  3.8 T:
the tracker, the electromagnetic and hadron calorimeters are within the field volume; while outside in the iron yoke, it is located the muon spectrometer for muon identification, momentum measurement and triggering. The $\mu$ spectrometer is made of six and four planes of Resistive Plate Chambers (RPC), in barrel and endcap region, respectively, arranged in four stations; the barrel is divided into 5 separate wheels  (named YB$\pm$2, YB$\pm$1 and  YB0) and 4 independent disks both in the positive and negative endcaps (named RE$\pm$4, RE$\pm$3, RE$\pm$2, RE$\pm$1). The 4th disk has been installed during the LHC long shutdown in  2013-2014 (LS1).  Three different gaseous detector technologies are employed: Drift Tube chambers (DT) in the barrel region to detect muons up to pseudo-rapidity |$\eta$|  < 1.2;  Cathode Strip Chambers (CSC) to handle the higher rates and non-uniform magnetic field in the endcap region 0.9 < |$\eta$| < 2.4; Resistive Plate Chambers (RPC) located in both barrel and endcap regions, up to |$\eta$| < 1.6. Each barrel wheel is divided into 12 sectors covering the full azimuthal angle ($\phi$); one sector consisting of four stations (named MB1 $\div$ 4) made by 4 layers of DT chambers and 6 layers of RPCs. In total there are  912 chambers, covering an area of 3500 m$^{2}$,  equipped with about 100.000 readout strips.   The CMS RPCs are double-gap chambers with 2 mm gas width each and copper readout in between;  operated in avalanche mode and with a 
bakelite bulk resistivity of  2 - 5 $\cdot$10$^{10}$ $\Omega $ cm; 
the gas mixture is made of 95.2\% C2H2F4, 4.5\% iC4H10 and 0.3\% SF6 with 40\% of humidity, working in closed loop mode  with about 10\% of fresh gas mixture  \cite{muon}. 

\section{2010-2012 RPC system operations} 

On 30th March 2010, LHC started proton-proton collisions at a center-of-mass energy of 7 TeV. 
 Commissioning of the machine through all the year turned into a substantial increase of the maximum instantaneous luminosity, achieving up to 2$\cdot$10$^{32}$ cm$^{-2} $s$^{-1}$  by the end of 2010. 2011 operations have seen enormous increase in luminosity, such that the entire 2010 dataset was delivered in less than a day. The instantaneous luminosity reached up to about 3.5$\cdot$10$^{33} $ cm$^{-2} $s$^{-1} $, setting a new world record for beam intensity at a hadron collider. In 2012, LHC  has been operating at the increased energy of 4 TeV per beam, reaching a maximum of instantaneous luminosity of about  6.6$\cdot$10$^{33}$ cm$^{-2}$s$^{-1}$. The total delivered luminosity was 30 fb$^{-1} $, out of this 27 fb$^{-1} $ and 25.2 fb$^{-1} $ were recored and validated by CMS, respectively.  The results discussed in the following sections are based on data collected in the period 2010-2012, called RUN1.    \\
During RUN1, the RPC system performed extremely well: the contribution to the CMS downtime was below 1.5\%, quite constant every year. Between 98\% and 97.5 \% of RPC channels were operational throughout RUN1 period. The fraction of dead (masked or inactive)  channels is mainly caused by noisy chambers due to faulty electronic boards located inside the chambers, not accessible since 2009, or  failures of HV/LV channels.   Most of inactive channels have been already recovered during LS1 \cite{anna_rpc2014}.

\section{RPC working point calibration}  

\begin{figure}[tb]
\includegraphics[width=0.52\textwidth] {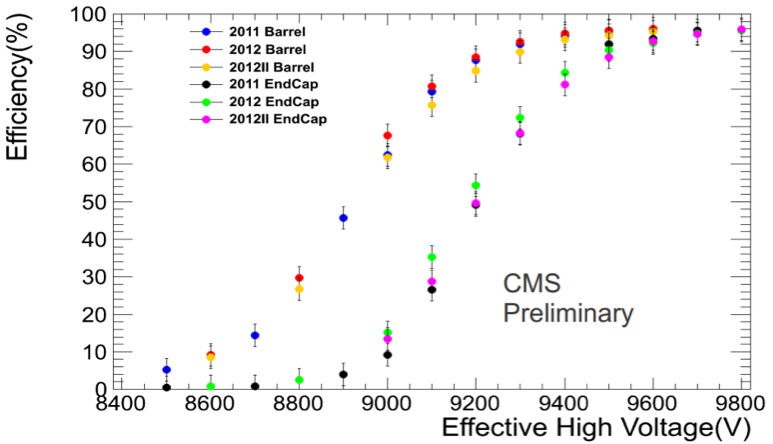}
\includegraphics[width=0.48\textwidth] {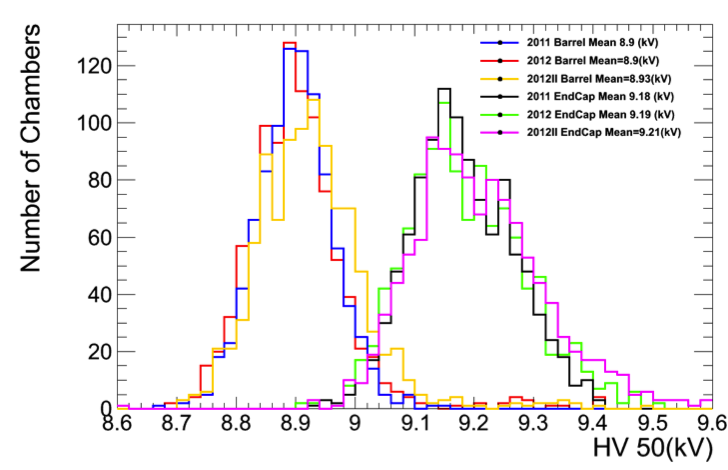}        
\caption{ Left: barrel and endcap average efficiency measured with the 2011 and 2012 HV 
        calibrations as a function of the HV$_{eff}$.  Right: the HV distributions  as measured at the 50\% of the efficiency.} 
    \label{eff_scan}
\end{figure}

During 2010, the LHC low luminosity did not allowed to tune the working voltage chamber by chamber. Thus
 only two different values were chosen to operate the endcap and barrel chambers  (9.5 kV and 9.3 kV, respectively). These values were defined on the basis of the measurements performed during the construction and commissioning phases. In early 2011 and  twice in 2012, High Voltage (HV) calibrations were carried out on collision data to optimize chamber Working Point (WP) and monitor in time the performance. The chamber WP  has been defined as the HV corresponding at the 95\% of the maximum efficiency plus 100 V for the barrel and 150 V for the endcap. 
 Muon events have been selected, thanks to the redundancy of the muon system, asking for a DT or CSC trigger. Then a linear extrapolation of track segment in DT and CSC chambers was performed toward the closest RPC strip plane and then matched to any RPC cluster in a range of 8 strips around the extrapolated impact point. More details on the method can be found in \cite{silvia}.   In order to take into account atmospheric pressure variations the HV was corrected for pressure variations defining the effective HV   (HV$_{eff}$) \cite{Pcorrection}.
 Fig.~\ref{eff_scan} shows, on the left, the average efficiency for all barrel and endcap chambers  as a function of HV$_{eff}$  measured in 2011 and 2012  and, on the right, the HV distributions as measured at 50\% of the efficiency.  No significative variations have been observed on both working voltage and chamber efficiency plateau values. The HV shift between barrel and endcap chambers depends on few $\mu$m difference in spacer width (located inside the gas gap).    
 
\section{RPC performance stability}
   
\begin{figure}[tbp]
\includegraphics[width=0.52\textwidth] {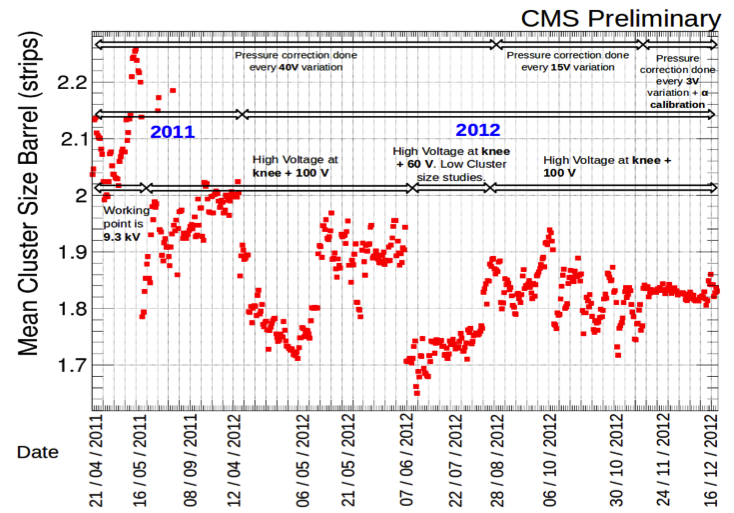}
\includegraphics[width=0.48\textwidth] {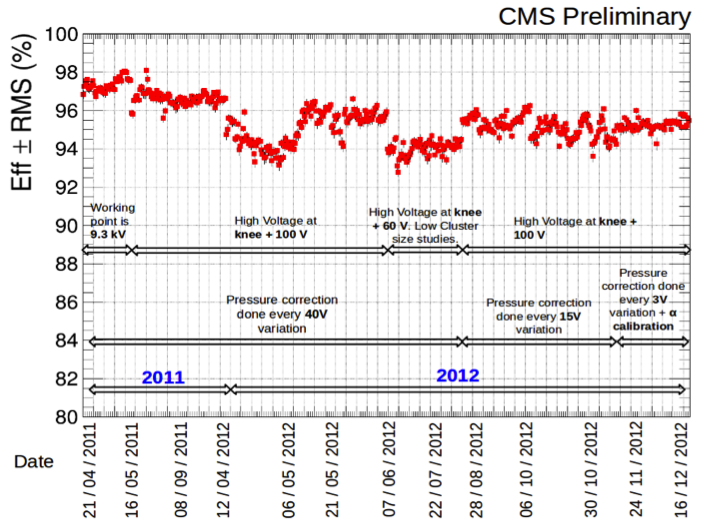}        
 \caption{Left: barrel cluster size vs. time. Right:  barrel average efficiency vs. time. The horizontal arrows separate runs taken applying  different  procedure for HV pressure correction.} 
    \label{monitor}
\end{figure}

During RUN1, the chamber performance was measured, run-by-run, in order to monitor the stability of the system. In Fig.~\ref{monitor} 
 the barrel average cluster size and efficiency as a function of time are shown. Different periods, distinguishable by the horizontal arrows, are highlighted in the plots, corresponding to different algorithms applied to correct the HV for the variation of the atmospheric pressure (the temperature of the caver was almost constant). At beginning, the WPs were corrected only when pressure variation effected a variation of the HV of 40 V;  then, at the end of July 2012, this value was reduced to 15 V and finally, in November 2012, to 3 V. In addition, during last period the pressure correction formula was modified with following formula, using $\alpha$ = 0.8: 
 \begin{equation}
HV = HV _{eff}(1-\alpha+\alpha p/p_{0})
\end{equation}
The improvement on detector stability is evident in both cluster size and efficiency monitoring plots. In particular, the efficiency variations of about 3\%, observed in the first period, were reduced to 1\% and then to less than $\pm$ 0.5 \% at the end. 
  
\begin{figure}[tbp]
\includegraphics[width=0.52\textwidth] {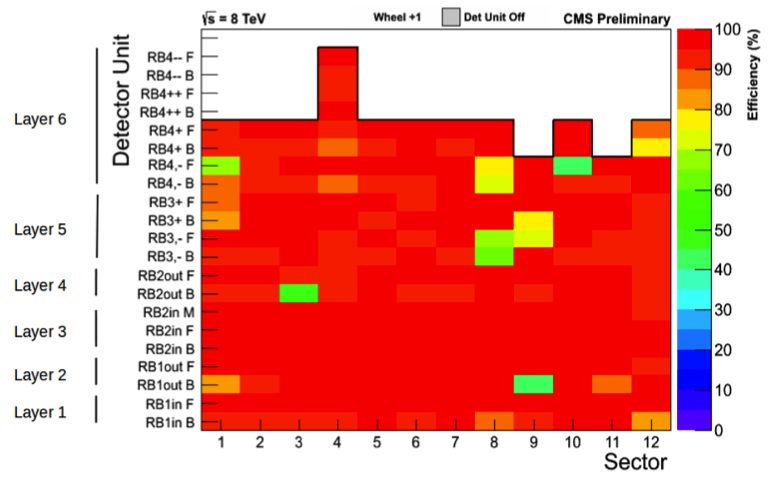}
\includegraphics[width=0.48\textwidth] {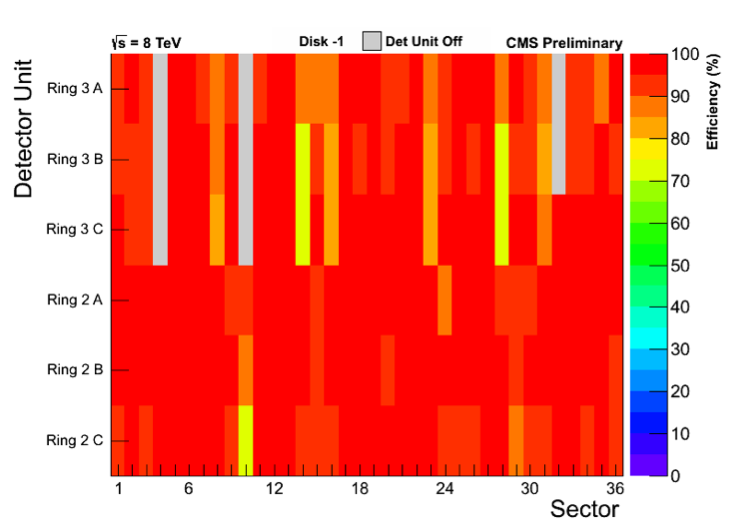}        
 \caption{ Left:  chambers efficiency map in one of barrel wheel +1. In $x$ axis the 12 sectors and in $y$ axis the 6 layers. Right: chambers efficiency map in one of endcap disk -1. In $x$ axis the 26 sectors and in $y$ axis the 3 layers.} 
    \label{eff_map}
\end{figure}

Fig.~\ref{eff_map}  shows the chambers efficiency map  as measured in 2012 for one wheel (YB+1) on the left and for one disk (RE-1) on the right. Most of the chambers have an efficiency more  than 95\%, except few cases where the lower efficiency is due to known hardware problems.

\section{2010 - 2012 noise and background rates}  

\begin{figure}[tbp]
\includegraphics[width=0.5\textwidth] {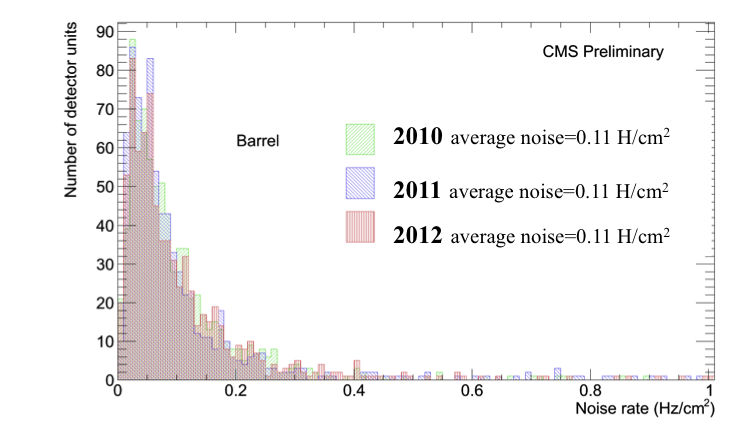}
\includegraphics[width=0.5\textwidth] {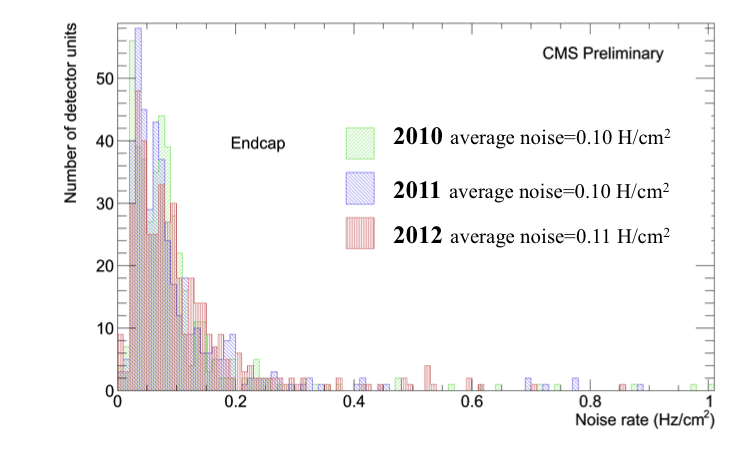}        
        \caption{Left: 2010 - 2012 noise rate distributions for barrel. Right:  2010 - 2012 noise rate distributions for endcap} 
    \label{noise}
\end{figure}

The intrinsic chamber noise and background radiation levels  can have important impact on the overall performance of the system: high values could affect the trigger performance and pattern recognition efficiency of muon tracks. 
Before every proton beam fill, the intrinsic noise rate is measured in order to mask noisy strips where needed.  Fig.~\ref{noise} shows the 2010-12 intrinsic noise distributions for barrel and endcap chambers on left and right plot, respectively. The average values, about 0.1 Hz/cm$^{2}$, are well below the CMS requirements and stable during the first three years of LHC. 

Background radiation in the muon chambers arises from  low-energy gamma rays and neutrons from p-p collisions,
low-momentum primary and secondary muons, 
punch-through hadrons from the calorimeters,
muons and other particles produced in the interaction of the beams with
collimators, residual gas and beam pipe elements.  

\begin{figure}[tbp]
\centering
\includegraphics[width=0.6\textwidth] {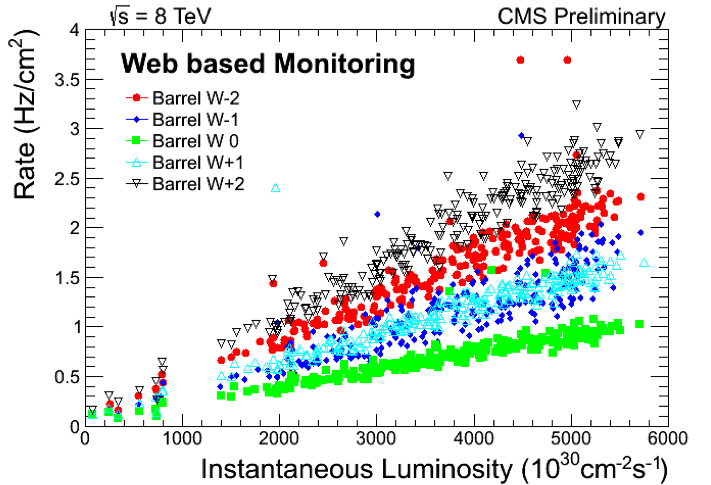}
        \caption{ 2012 measured rate in barrel wheels as a function of the LHC luminosity (YB0 is the wheel closest to the interaction point, YB+2 and -2 are farthest way).}
    \label{rate_lin}
\end{figure}

Fig.~\ref{rate_lin}  shows the average rates versus the instantaneous 2012 LHC luminosity as measured in each five barrel wheels. 
As expected the  rates increase as the chambers are farther from the IP: YB0 is the wheel closest to the interaction point, YB+2 and -2 are farthest way.  

\begin{figure}[tbp]
\centering
\includegraphics[width=0.47\textwidth] {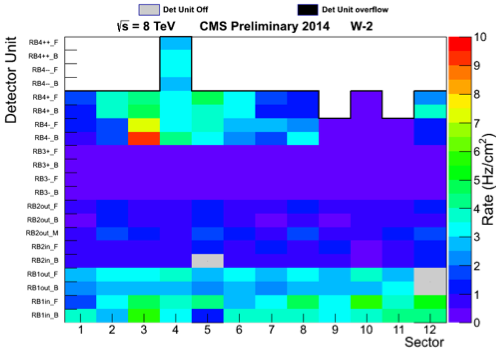}
\includegraphics[width=0.47\textwidth] {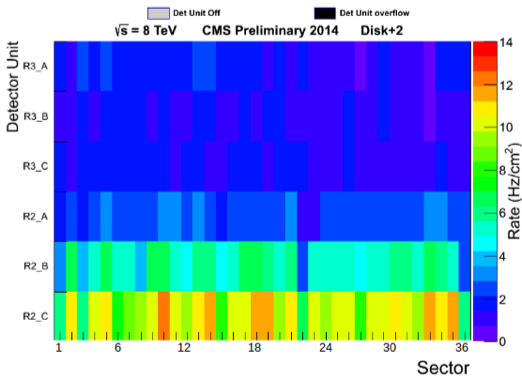}        
 \caption{Left:  measured noise rate map in one of the barrel wheel -2.  Right:  measured noise rate map in one of the endcap disk +2. LHC luminosity about $ 4x10^{33}$ cm$^{-2}$s$^{-1}$} 
    \label{background}
\end{figure}

Fig.~\ref{background} shows  the chambers rate map as measured in one  barrel wheel (on left) and one endcap disk (on right) at a luminosity of $ 4x10^{33}$ cm$^{-2}$s$^{-1}$.  In the barrel region, the highest rates ($\sim$~7~Hz/cm$^{2}$) are in the outermost stations (RB4), mainly due to diffused neutron background, and  in the innermost stations (RB1), mainly due to charged particles and punch trough hadrons; whereas in the endcap region the highest rate ($\sim$~12 ~Hz/cm$^{2}$) is measured in disk 2, inner ring. More details can be found  \cite{silvia_rpc2014}.

At the LHC and HL-LHC center-of-mass energy of 14 TeV and peak luminosity of $2x10^{34}$ cm$^{-2}$s$^{-1}$  and  $5x10^{34}$ cm$^{-2}$s$^{-1}$, respectively,  the rates are expected to extrapolate linearly compared to the current data. 
Table~\ref{expected_rate} summarizes the maximum and average expected background rates.  
Such high radiation background will push RPC system beyond its original lifetime  \cite{aging1} \cite{aging2}, thus requiring a new assessment of the detector longevity (radiation tolerance, aging of components, long-term 
behavior) and performance  under accelerated aging tests. 

\begin{table}[ht!]
\caption{Expected maximum and average background conditions  at  LHC luminosity of $2 x 10^{34}$ cm$^{-2}$s$^{-1}$  (for phase 1) and at HL-LHC  luminosity of $5x10^{34}$ cm$^{-2}$s$^{-1}$. No safety factor is included in the estimation.}
\label{expected_rate}
\smallskip
\centering
\begin{tabular}{c|  c  c | c c  } 
 &  \multicolumn{2}{c}{ \bf{L = $2x 10^{34}$ cm$^{-2}$s$^{-1}$}} & \multicolumn{2}{c}{\bf{L= $5x 10^{34}$ cm$^{-2}$s$^{-1}$ }} \\ 
\hspace{1cm}
& Max. rate (Hz/cm$^{2}$) & <Rate> (Hz/cm$^{2}$) & Max. rate (Hz/cm$^{2}$) & <Rate> (Hz/cm$^{2}$)\\ 
\hline 
Barrel & 50 & 10 & 125 & 25  \\
Endcap & 100 & 25 & 250 & 60 \\
\end{tabular}
\end{table}

\section{Conclusion}   
The operation of the CMS RPC muon system during the first three years of LHC data taking was outstanding:  the contribution to CMS downtime was below 1.5 \% and, at the end of RUN1, the fraction of active channels was about 97.5 \%. 
Most of inactive channels have been already recovered during LS1. 
After 3 years of LHC running with increasing instantaneous luminosity and 6 years from the end of RPC construction, 
the detector performance is within the CMS specifications and stable with no degradation observed. 
The monitoring in time of the RPC performance shows an average efficiency above 95\%, stable within $\pm$ 0.5 \% of variations. The expected background rate at HL- LHC luminosity of 5 $x 10^{34} $cm$^{-2} $s$^{-1} $ will be sustainable for the present RPC system, even if new aging tests are needed to certified the system to 10 HL - LHC years.

\end{document}